\documentstyle[preprint,aps]{revtex}

\begin{document}


\title{Transition from anomalous to normal hysteresis 
in a system of coupled Brownian motors: a mean field approach}

\author{S. E. Mangioni and R. R. Deza}

\address{Departamento de F\'{\i}sica, FCEyN, Universidad Nacional de
Mar del Plata\\ De\'an Funes 3350, 7600 Mar del Plata, Argentina.}

\author{H. S. Wio}

\address{Centro At\'omico Bariloche (CNEA) and Instituto Balseiro
(CNEA and UNCuyo)\\ 8400 San Carlos de Bariloche, Argentina.}

\date{\today}

\maketitle

\begin{abstract}
We address a recently introduced model describing a system of 
periodically coupled nonlinear phase oscillators submitted to {\em 
multiplicative\/} white noises, wherein a ratchet-like transport 
mechanism arises through a symmetry-breaking noise-induced 
nonequilibrium phase transition.  Numerical simulations of this system 
reveal amazing novel features such as {\em negative zero-bias 
conductance\/} and {\em anomalous hysteresis}, explained 
resorting to a strong-coupling analysis in the thermodynamic 
limit.  Using an explicit mean-field approximation we explore the 
whole ordered phase finding a transition from anomalous to normal 
hysteresis inside this phase, estimating its locus and identifying 
(within this scheme) a mechanism whereby it takes place.
\end{abstract}

\section{Introduction}\label{sec.1}

Feynman's ratchet-and-pawl example \cite{[1]}, illustrating the
impossibility for a microscopic rectifying device to extract work in a
cyclic manner from the \emph{equilibrium} fluctuations of a
\emph{single} heat bath, spurred in turn the search for heat engines
operating in a \emph{far from equilibrium} regime between \emph{two}
heat baths.  The field of ``nanomechanics'' (more
specifically, that of noise-induced transport or ``Brownian motors'')
is now about one decade old \cite{[3]}.  In the early works, a requisite
for these devices to operate (besides their obvious built-in,
ratchet-like, bias) seemed to be that the fluctuations be correlated
\cite{[4]}.  That requirement was relaxed when ``pulsating'' ratchets
were discovered: in these it is the random \emph{switching} between
uncorrelated noise sources which is responsible of the rectifying effect
\cite{[5]}.

A recent new twist has been to relax also the requirement 
of a built-in bias \cite{[6]}: a system of periodically coupled nonlinear
phase oscillators in a symmetric ``pulsating'' environment has been
shown to undergo a noise-induced nonequilibrium phase transition,
wherein the spontaneous symmetry breakdown of the stationary
probability distribution gives rise to an {\em effective}
ratchet-like potential.  The authors introduced the
aforementioned mechanism and its striking consequences, such as the
appearance of {\em negative zero-bias conductance} and 
{\em anomalous hysteresis}, which they illustrated through numerical
simulations and explained by resort to the strong-coupling limit. 
By anomalous hysteresis we refer to the case where the cycle runs 
clockwise, in opposition to the normal one (as typified by a ferromagnet) 
that runs counterclockwise. 

Exploiting our previous experience in a lattice model displaying (like 
the present one) a symmetry-breaking nonequilibrium phase transition 
\cite{[8]} and in order to set a firm ground for further work, 
we addressed the model using an explicit mean-field approach 
\cite{[7]}, focusing on the relationship between the \emph{shape} of 
the stationary probability distribution (as well as the \emph{number 
of solutions} to the mean-field equations) and the transport 
properties in its different regions.  Hence it is our aim in this work 
to report on the thorough exploration of the ordered phase, on the 
characterization of its subregions, and on features related to the 
transition from anomalous to normal hysteresis in the behavior of the 
particle current as a function of the bias force.  Our main finding is 
that there exists a close relationship between the character of the 
hysteresis loop on one hand, and the shape of the stationary 
probability distribution as well as the number of ``homogeneous'' 
solutions (a term to be clarified later) on the other.

In the following sections we successively introduce the model,
describe the mean-field approach, discuss our numerical results and
draw our conclusions.  For the benefit of the reader we
have included an appendix where some (in our judgment) subtle
calculations are performed in some detail.

\section{The model}\label{sec.2}              

In Ref.\cite{[6]} the authors consider a set of {\em globally\/}
coupled stochastic equations of motion (in the overdamped regime, and
to be interpreted in the sense of Stratonovich) for the $N$ degrees of
freedom (phases) $X_i(t)$:
\begin{equation}
        \dot X_i=-\frac{\partial U_i}{\partial X_i}+\sqrt{2T}\,
        \xi_i(t)-\frac{1}{N}\sum_{j=1}^N K(X_i-X_j).\label{eq:1}
\end{equation}
The model just set up can be visualized (at least for some parameter values) 
as a set of overdamped interacting pendula. The second term models, as usual, 
the effect of thermal fluctuations: $T$ is the temperature of the environment 
and the $\xi_i(t)$ are {\em additive\/} Gaussian white noises with zero mean 
and variance one
\begin{equation}
        \langle\xi_i(t)\rangle=0,\qquad\langle\xi_i(t)\xi_j(t')\rangle=
        \delta_{ij}\delta(t-t').\label{eq:2}
\end{equation}

The ``pulsating'' potentials $U_i(x,t)$ are among the key ingredients
in the model \cite{[5]} ($x\in[-L/2,L/2]$ is a phaselike real variable
that runs over the range of $X_i(t)$, namely the allowed values for
any realization of any $X_i$ at any time $t$). They consist of a
static part $V(x)$ and a fluctuating one: Gaussian white noises
$\eta_i(t)$ with zero mean and variance one (i.e.\ obeying also Eq.\
(\ref{eq:2}) but assumed \emph{nonthermal} in origin) are coupled {\em
multiplicatively\/} (with intensity $Q$) through a function $W(x)$.
Even though we adopted $L=2 \pi$ in all our numerical calculations,  
we kept in all the equations the dependence on $L$ in order to have 
the most general expressions. 
For the analysis of the noise-induced ratchet effect a ``load force''
$F$, producing an additional bias, is also included 
\begin{equation}
        U_i(x,t)=V(x)+W(x)\sqrt{2Q}\,\eta_i(t)-Fx.\label{eq:3}
\end{equation}
As already stated, both $V(x)$ and $W(x)$ are assumed to be {\em
periodic\/} (period $L$) and moreover they are {\em symmetric\/}:
\begin{displaymath}
        V(-x)=V(x)\mbox{ , }W(-x)=W(x)
\end{displaymath}
which means that there is no \emph{built-in} ratchet effect.  In
Ref.\cite{[6]} the choice was 
\begin{equation}
        V(x)=W(x)=-\cos x-A\cos{2x},\label{eq:4}
\end{equation}
with $A>0$ so as to remove an accidental degeneracy hindering the 
spontaneous symmetry breakdown \cite{[6]}, but not strong enough to create 
a local minimum at $L/2=\pi$.  With the choice $A>0$ the direction of 
the particle current $\langle\dot X\rangle$ turns out to be 
\emph{opposite} to that of symmetry breaking in the stationary 
probability distribution $P^{st}(x)$, and it is this effect which 
leads in turn to such oddities as {\em negative zero-bias 
conductance\/} and {\em anomalous hysteresis} \cite{[6]} (a nice 
animation illustrating this phenomenon can be found on the web \cite{[10]}).

The interaction force $K(x-y)=-K(y-x)$ between oscillators (the other
key ingredient in the model) is also assumed to be a {\em periodic\/}
function of $x-y$ with the same period $L=2\pi$ as $V(x)$, $W(x)$.  In
Ref.\cite{[6]} it is
\begin{equation}
        K(x)=K_0\sin x,\qquad K_0>0.\label{eq:5}
\end{equation}
As indicated before, the model can be visualized (at least for $A\to 0$) 
as a set of overdamped and interacting pendula (only their phases matter, 
not their locations) interacting with one another through a force proportional
to the sine of their phase difference (a force that is always
attractive in the reduced interval $-\pi\le x-y\le\pi$).

Summarizing, the complete set of parameters in the model is: $N, F, T, A,
K_0$ and $Q$.  Except for numerical simulations the exact value of $N$ 
is unimportant, as long as it is large.  As already said, $F$ is just 
an auxiliary tool
for the analysis.  We shall fix the values of $T$ and $A$ as in
Ref.\cite{[6]}, namely $T=2, A=0.15$.  So the important parameters in
the model are $K_0$ (governing mostly the ``drift'' terms in this set
of generalized Langevin equations) and $Q$ (governing mostly the
``diffusion'' ones).  As discussed in Ref.\cite{[5]}, the Gaussian
character of $\eta_i(t)$ allows it to be added to $\xi_i(t)$.  Hence
it suffices to consider the $\eta_i(t)$'s, now coupled through
$S(x)\equiv\sqrt{2}[T+Q(W')^2]^{1/2}$ (note that $S(x)=\sqrt{2}g(x)$
as defined in Ref.\cite{[6]}).

\section{Mean-field analysis}\label{sec.3}

On account of the choice made in Eq.\ (\ref{eq:4}), the interparticle
interaction term in Eq.\ (\ref{eq:1}) can be cast in the form:
\begin{equation}
        \frac{1}{N}\sum_{j=1}^N K(X_i-X_j)=
        K_0\left[C_i(t)\sin X_i-S_i(t)\cos X_i\right].\label{eq:6}
\end{equation}
For $N\to\infty$ we may approximate Eq.\ (\ref{eq:1}) \`a la
Curie-Weiss, replacing $C_i(t)\equiv N^{-1}\sum_j\cos x_j(t)$ and
$S_i(t)\equiv N^{-1} \sum_j\sin x_j(t)$ by $C_m\equiv\langle\cos
x_j\rangle$ and $S_m\equiv\langle\sin x_j\rangle$ respectively, to be
determined as usual by self-consistency.  This decouples the system of
stochastic differential equations (SDE) in Eq.\ (\ref{eq:1}), which
reduces to essentially one Markovian SDE for the single stochastic
process $X(t)$:
\begin{equation}
        \dot X=R(X)+S(X)\eta(t),\label{eq:7}
\end{equation}
with
\begin{eqnarray}
        R(x)&=&-V'(x)+F-K_m(x)\nonumber\\
        &=&-\sin x(1+K_0C_m+4A\cos x)+K_0S_m\cos x+F\label{eq:8}
\end{eqnarray}
(where $K_m(x)=K_0[C_m\sin x-S_m\cos x]$) and
\begin{equation}
        S(x)=\sqrt{2\{T+Q[W'(x)]^2\}}=\sqrt{2\{T+Q[\sin x+2A\sin 2x]^2\}},
        \label{eq:9}
\end{equation}
so that the ``spurious'' contribution to the drift in the Stratonovich
interpretation is
\begin{equation}
        \frac{1}{2}S(x)\,S'(x)=QW'(x)W''(x)=Q(\sin x+2A\sin 2x)(\cos x+
        4A\cos 2x).\label{eq:10}
\end{equation}
We may write $R(x)=-{\cal V}'(x)$ in terms of an \emph{effective}
(since $C_m$ and $S_m$ are determined by self-consistency in terms of
$P^{st}(x)$ below and moreover by the asymmetric, hence ratchet-like)
potential
\begin{eqnarray*}
        {\cal V}(x)&=&V(x)+\int_{0}^{x}K_m(y)dy-F\,x\\
        &=&-[\cos x(1+K_0C_m)+A\cos 2x]-K_0S_m\sin x-F\,x.
\end{eqnarray*}
Similarly, $S(x)S'(x)/2$ can be derived from $-Q[W'(x)]^2/2$.

\subsection{The stationary probability distribution function}\label{subsec.a}

The Fokker-Planck equation associated with the SDE in Eq.\ (\ref{eq:7}) is
\begin{equation}
        \partial_t P(x,t)=\partial_x\{-[R(x)+\frac{1}{2}S(x)\,S'(x)]P(x,t)
        \}+\frac{1}{2}\partial_{xx}[S^2(x)P(x,t)]\label{eq:11}
\end{equation}
(see the appendix, Eqs.\ (\ref{eq:A3}) and (\ref{eq:A4})) and its
normalized stationary solution \emph{with periodic boundary
conditions} and current density $J\ne 0$ is \cite{[5],[6]}
\begin{equation}
        P^{st}(x)=\frac{e^{-\phi(x)}\,H(x)}{{\cal N}\,S(x)},
        \label{eq:12}
\end{equation}
where
\begin{equation}
        \phi(x)=-2\int_0^x dy\,\frac{R(y)}{S^2(y)},\label{eq:13}
\end{equation}
\begin{equation}
    H(x)=\int_{x}^{x+L}dy\,\frac{\exp[\phi(y)]}{S(y)},\label{eq:14}
\end{equation}
and ${\cal N}=\int_{-L/2}^{L/2}dx\,P^{st}(x)$.  The positive sign of
$S(x)$ and the exponentials implies that of $H(x)$ and hence that of
$P^{st}(x)$ and ${\cal N}$, as it should be.  On the other hand,
although $R(x)$, $S(x)$ and $P^{st}(x)$ are periodic by construction,
$\phi(x)$ is not required to be so; in fact it increases on each cycle
by an amount
\begin{equation}
        \phi(L)=\frac{1}{T}\int_{0}^{L}\frac{dx\,p(x)\sin x}{q(x)}-
        \frac{K_0S_m}{T}\int_{0}^{L}\frac{dx\cos x}{q(x)}-
        \frac{F}{T}\int_{0}^{L}\frac{dx}{q(x)},\label{eq:22}
\end{equation}
with $p(x)=1+K_0C_m+4A\cos x$ and $q(x)=1+(Q/T)\sin^2 x\,(1+4A\cos
x)^2$, both even functions of $x$.  Since only the first term vanishes
identically, for nonzero $F$ or $S_m$ it will be the case 
\emph{generically} that $\phi(L)\neq 0$; thus the form of $H(x)$ in 
Eq. (\ref{eq:14}) is designed to compensate for this fact.  Moreover 
(as shown in the appendix) for $A>0$, $[q(x)]^{-1}(>0)$ gives less 
weight to the positive $\cos x$ values than to the negative ones, 
and it does increasingly so the larger $Q$ is; hence the two nonzero 
contributions to $\phi(L)$ compete with each other.

According to Eq.\ (\ref{eq:A15}) in the appendix 
\begin{equation}
        J=[1-e^{\phi(L)}]/2{\cal N},
        \label{eq:21}
\end{equation}
hence the sign of $J$ is that of $1-e^{\phi(L)}$ and\textemdash on the
other hand\textemdash the ``holonomy'' condition $e^{\phi(L)}=1$
implies $J=0$ and $H(x)=\mathrm{const.}$ (see the appendix, Eq.\
(\ref{eq:A8})).  Equation (\ref{eq:21}) is a self-consistency relation
since both ${\cal N}$ and $\phi(L)$ keep information on the shape of
$P^{st}(x)$ (in the latter case through $C_m$ and $S_m$).  A nonzero
$J$ is always associated with a symmetry breakdown in $P^{st}(x)$
(namely, $P^{st}(-x)\neq P^{st}(x)$).  This may be either
\emph{spontaneous} (our main concern here) or \emph{induced} by a
nonzero $F$.

\subsection{The self-consistency equations}\label{subsec.b}

As indicated earlier, the stationary probability distribution $P^{st}(x)$
depends on both $S_m$ and $C_m$, since $R(x)$ includes $K_m(x)$.
Their values arise from requiring self-consistency, which amounts to
solving the following system of nonlinear integral equations 
\begin{eqnarray}
        F_{cm}&=&C_m,\mbox{  with  }F_{cm}\equiv\langle\cos x\rangle=
        \int_{-L/2}^{L/2}dx\,\cos x\,P^{st}(x,C_m,S_m),\label{eq:15}\\
        F_{sm}&=&S_m,\mbox{  with  }F_{sm}\equiv\langle\sin x\rangle=
        \int_{-L/2}^{L/2}dx\,\sin x\,P^{st}(x,C_m,S_m).\label{eq:16}
\end{eqnarray}
These equations give $C_m$ and $S_m$ for each set of the parameters
($Q$, $K_0$) that define the state of the system (assuming $T$, $A$ and
$F$ fixed).

For $F=0$, the choice $S_m=0$ makes $R(x)$ in Eq.\ (\ref{eq:8}) an odd
function of $x$; this in turn makes $\phi(x)$ in Eq.\ (\ref{eq:13})
even, and then the periodicity of $P^{st}(x)$ in Eq.\ (\ref{eq:12})
(in the form $P^{st}(-x)=P^{st}(-x-L)$) implies that the
stationary probability distribution is also an even function of $x$.  So
the problem of self-consistency reduces to the \emph{numerical} search
of solutions of Eq.\ (\ref{eq:15}), with $S_m=0$.  Nonetheless, a
plausibility argument leads to an intuition on the existence of
some solutions of this integral equation (and their stability) in this
symmetric case:
\begin{itemize}
        \item Since\textemdash as argued\textemdash  $J=0$ and
        $H(x)=\mathrm{const.}$ holds, it turns out from Eq.\ (\ref{eq:11}) 
        that the set of critical points $x_{c}$ of $P^{st}(x)$ must obey
        $R(x_{c})=\frac{1}{2}S(x_{c})\,S'(x_{c})$.

        \item Since $x_{c}=0,\pi\,(\mathrm{mod}\:2\pi)$ belong to that
        set, a possible way to satisfy the integral equation in 
        Eq. (\ref{eq:15}) is for $P^{st}(x)$ be concentrated 
        around those values.

        \item Then, by analogy with a pendulum, one expects the solution 
        with $C_m>0$ to be the stable one for $K_0/Q$ large enough 
        (what we shall call an ``interaction-driven regime'' or 
        \emph{idr}).  For $K_0/Q$ small enough (``noise-driven 
        regime'' or \emph{ndr}) the stable solution \emph{can} have 
        $C_m<0$ (implying an ``angle'' larger than $\pi/2$ with 
        respect to the mean field) since it corresponds to shaking the 
        pendula violently.
\end{itemize}
Beyond these handwaving arguments it must be said that, since $\cos x$
in Eq.\ (\ref{eq:15}) is an even function of $x$, in order to determine
the stability of the true solutions it suffices to use the Curie-Weiss
(i.e., the one-parameter) criterion, namely to check whether the slope
at $S_m$ of the integral in Eq.\ (\ref{eq:16}) is less or greater than
one.  As a complementary check, a small-$x$ expansion of $\phi(x)$
(see the appendix) confirms that $P^{st}(x)$ is indeed Gaussian at
$x=0$.  For \emph{small} $F\neq 0$, $P^{st}(x)$ gets multiplied (in
this approximation) by $\exp{[Fx/T]}(\approx 1+Fx/T)$ which leads to a 
nonzero value of $S_m=kF$, with $k>0$.  By the mechanism discussed after 
Eq.\ (\ref{eq:22}), for $Q$ large enough  $\phi(L)>0$ and by 
Eq.\ (\ref{eq:21})  $J<0$.  As will be shown later, this effect 
manifests itself in a {\em negative zero-bias conductance\/}. 
%
%

We conclude that, for $F=0$ there are always one or more solutions to
Eqs.\ (\ref{eq:15})\textendash(\ref{eq:16}) with $S_m=0$ and one of
these is the stable one in the ``disordered'' phase.  As argued in
Ref.\cite{[6]}, for $N\to\infty$ a noise-induced nonequilibrium
transition takes place \emph{generically} towards an ``ordered'' phase
where $P^{st}(-x)\neq P^{st}(x)$.  In the present scheme this
asymmetry should be evidenced by the fact that the solution with
$S_m=0$ becomes unstable in favor of other two solutions such that
$P^{st}_{2}(x)=P^{st}_{1}(-x)$, characterized by \emph{nonzero} values
$\pm|S_m|$.  This fact confers $S_m$ the character of an order parameter.

\subsection{The phase boundary}\label{subsec.c}

Since $\sin x$ is an antisymmetric function, Eq.\ (\ref{eq:16}) results
impractical for the task of finding the curve that separates the
ordered phase from the disordered one, given that on that curve $S_m$
is still zero.  For that goal (exclusively) we solve, instead of Eqs.\
(\ref{eq:15})--(\ref{eq:16}), the following system:
\begin{eqnarray}
        \int_{-L/2}^{L/2}dx\,\cos x\,P^{st}(x,C_m,0)&=&C_m,\label{eq:17}\\
        \int_{-L/2}^{L/2}dx\,\sin x\,\left.\frac{\partial P^{st}}
        {\partial_{S_m}}\right|_{S_m=0}&=&1.\label{eq:18}
\end{eqnarray}

\subsection{The particle current}\label{subsec.d}

The appearance of a ratchet effect amounts to the existence of a
nonvanishing drift term $\langle\dot X\rangle$ in the stationary
state, in the absence of any forcing ($F=0$); in other words, the
pendula become rotators in an average sense.  As discussed above,
the cause of this spontaneous particle current is the noise-induced
asymmetry in $P^{st}(x)$ \cite{[6]}.

As it is shown in the appendix
\begin{equation}
    \langle\dot X\rangle=
        \int_{-L/2}^{L/2}dx\left[R(x)+\frac{1}{2}S(x)S'(x)\right]
        P^{st}(x,C_m,S_m),\label{eq:19}
\end{equation}
and the final result is
\begin{equation}
        \langle\dot X\rangle=J\,L=\left\{\frac{1-e^{\phi(L)}}{2{\cal N}}
        \right\}L.\label{eq:20}
\end{equation}
Hence $\langle\dot X\rangle$ has the sign of $J$ and can be also
regarded as an order parameter.  In fact, from Eqs.\ (\ref{eq:8}) and
(\ref{eq:19})\textemdash or equivalently from Eqs.\ (\ref{eq:8}),
(\ref{eq:13}) and (\ref{eq:20})\textemdash one may suspect the
existence of a tight relationship between $\langle\dot X\rangle$ and
$S_m$.

\section{Numerical results}\label{sec.4}

Figure \ref{fig:1} displays (on the same scale as Fig.\ 1b of
Ref.\cite{[6]}, with which it fully coincides) the phase diagram
obtained by solving Eqs.\ (\ref{eq:17})--(\ref{eq:18}) using the
Newton-Raphson method.  In the region above the full line (``ordered
region'') the stable solution to Eqs.\ (\ref{eq:15})--(\ref{eq:16})
has $S_m\neq 0$.  Notice that this noise-induced phase transition is
\emph{reentrant}: as $Q$ increases for $K_0=\mathrm{const.}$, the
``disordered phase'' ($S_m=0$) is met again.  The multiplicity of
mean-field solutions in the ordered region, together with the fact
that some of them may suddenly disappear as either $K_0$ or $Q$ are
varied (a fact that, as we shall see, is closely related to the
occurrence of anomalous hysteresis) hinder the pick of the right
solution in this region.

A more systematic characterization of the aforementioned multiple 
solutions is achieved when the branch to which they belong is followed 
from its corresponding ``homogeneous'' ($S_m=0$) solution.  
Accordingly, the dashed line in Fig.\ \ref{fig:1} separates two 
sectors within the ordered region with regard to the 
\emph{homogeneous} solutions.  Below it (``noise-driven regime'' or 
\emph{ndr}) there is \emph{a single} solution with $S_m=0$ and $C_m<0$ 
(as already suggested, in this regime a solution with $C_m<0$ 
\emph{can} be stable since it corresponds to shaking violently the 
pendula).  Above it (``interaction-driven regime'' or \emph{idr}) 
there are \emph{three}: two of them have opposite signs and (for 
$K_0/Q$ large enough) $|C_m|\simeq 0.9$; the remaining one has 
$C_m\approx 0$.  Note that this line presents a dip, whose meaning
will be discussed in relation with the character of the hysteresis 
loop.  We have studied the shape of $P^{st}(x)$ and the behavior of 
$\langle\dot X\rangle$ as a function of $F$ for different locations in 
this $(Q,K_0)$ diagram.  The squares in Fig.\ \ref{fig:1} indicate 
several positions inside and outside the ordered zone for $K_0=10$.  
For this value, the separatrix between both regimes lies around $Q=6$.  
Figure \ref{fig:2} illustrates the crossing of the dashed line 
in Fig. 1, and the persistence of the negative solution after the 
disappearance of the other two.

\subsection{Analysis at constant coupling}\label{subsec.e}

Figure \ref{fig:3} shows (for the \emph{true} solution, namely the
\emph{stable} $S_m\neq 0$ one) the evolution of $P^{st}(x)$ as a
function of $Q$, for $K_0=10$: whereas for $Q=1$ and $Q=21$ it is a
\emph{symmetric} function of $x$, there is a spontaneous breakdown of
this symmetry for the remaining values (the system has to choose
between two possible \emph{asymmetric} solutions, of which just one is
shown).  A noticeable feature of $P^{st}(x)$ is that for some $Q>6$ it
becomes \emph{bimodal} (in fact, already for $Q=6$ do we see an
indication that a second peak is developing).  According to Fig.\
\ref{fig:3} and Eq.\ (\ref{eq:15}), for $Q<6$ it may be expected that
(for the \emph{true} solution) $C_m$ be positive and even relatively
large (however, between $Q=2$ and $Q=6$ other solutions are possible,
which lead to other shapes of $P^{st}(x)$ not shown).  As the second
peak develops and becomes higher than the original one, $C_m$ shifts
toward small negative values (in this region, the one depicted is the
only possible solution).

The squares in Fig.\ \ref{fig:4} plot (always for $K_0=10$) the two 
possible values of the spontaneous drift velocity $\left.\langle\dot 
X\rangle\right|_{F=0}$ in the ordered phase as functions of $Q$.  The 
vertical thick line indicates the value of $Q$ at which the transition 
from anomalous to normal hysteresis occurs for $K_0=10$.  The effect 
of a moderate positive bias $F$ on $\langle\dot X\rangle$ in the 
normal region (at the right of the thick line) is clearly 
understandable (the only surprising feature is that on the reentrant 
branch of the phase boundary the transition for $F\neq 0$ is so steep 
that it resembles a first order one).  But the most striking feature is 
the sudden disappearance of a ``forward'' particle current for the 
shown values of $F$ as we cross the thick line towards the left 
(although it may still exist for lower values of $F$): this is the 
manifestation of the anomalous hysteresis.  As suggested in the first 
paragraph of this section, this phenomenon is intimately related to 
the sudden disappearance of some of the multiple solutions when either 
$K_0$ or $Q$ are varied in the \emph{idr} (see Fig.\ \ref{fig:2}).

Figures \ref{fig:5}(a) to \ref{fig:5}(d) present a sequence of
$\langle\dot X\rangle$ vs $F$ plots, varying $Q$ across the thick line
of Fig.\ \ref{fig:4}.  In these, all the solutions to Eqs.\
(\ref{eq:15})--(\ref{eq:16}) but the one belonging to the branch
starting at $C_m\approx 0$ for $S_m=0$ have been included.  For
$Q=5.97$ (Fig.\ \ref{fig:5}(a)) two (unstable) solutions meet at
$\langle\dot X\rangle=0$ for $F=0$.  The progressive withdrawal of one
of them out of the $F\approx 0$ region with increasing $Q$ until its
complete disappearance (figs.\ \ref{fig:5}(b) to \ref{fig:5}(d)) can
be traced back (through their corresponding branches) to the
disappearance of solutions for $S_m=0$.  Moreover, it is only after
this solution has completely disappeared that the stable solution
begins to exist for larger values of $F$ and thus normal hysteresis
sets in (Fig.\ \ref{fig:5}(d)).

It is also instructive to see how the different branches in figs.\
\ref{fig:5}(a) to \ref{fig:5}(c) develop as one enters the ordered
region from the left.  Figure \ref{fig:6} shows for three points on
the $K_0=10$ line the $\langle\dot X\rangle$ vs $F$ plots for all the
existing solutions (but not the one belonging to the branch starting at
$C_m\approx 0$ for $S_m=0$).  For $Q=1$ there is a single stable
solution displaying negative zero-bias conductance; for $Q=1.7$ (right
on the phase boundary) a second (unstable) solution appears and the
anomalous hysteretic behavior (clearly seen for $Q=3$) sets in.  As
suggested by Fig.\ \ref{fig:4}, the situation is different at the
reentry: Figure \ref{fig:7} shows that the disappearance of the
(normal) hysteretic behavior at the phase boundary in the \emph{ndr}
is in fact abrupt, signalling a \emph{first order} phase transition in
this regime.

\subsection{Analysis at constant noise intensity}\label{subsec.f}

A complementary view of the transition in figs.\ \ref{fig:5}(a) to
\ref{fig:5}(d) is obtained by varying $K_0$ at $Q=6.0$ (figs.\
\ref{fig:8}(a) and \ref{fig:8}(b)): for $K_0=7.25$ a very small normal
hysteresis loop can be appreciated, which has grown rather large
already for $K_0=8.0$; for $K_0=10.0$ the loop has become anomalous
and a third branch has appeared, forming a cusp at the endpoints of
the loop; for a larger $K_0$ (Fig.\ \ref{fig:8}(b)) the cusp develops
into a curl.  For larger values of $Q$ (always across the dashed line
in Fig.\ \ref{fig:1}) the general pattern is about the same (see Fig.\
\ref{fig:9} for $Q=9.5$, where the kink of the normal loop at a
position rather close to the endpoint of the anomalous one is
suggestive); a similar plot to that in Fig.\ \ref{fig:2} (but now
varying $K_0$ at $Q=10.0$) is shown in Fig.\ \ref{fig:10}, where the
remaining solution displays a larger value of $|C_m|$.  Finally, it is
interesting to elucidate the nature of the transition \emph{at the
left} of the dip in the dashed line in Fig.\ \ref{fig:1}: as Fig.\
\ref{fig:11} shows, here the loss of two solutions is not accompanied
by a change in the character of the hysteresis loop, which remains
anomalous.

\section{Conclusions}\label{sec.5}

We have shown the existence of a sharp transition in the behavior of
the system inside the ordered phase, from an ``interaction-driven
regime'' (\emph{idr}) (typically for $K_0/Q$ larger than about three
halves) towards a ``noise-driven regime'' (\emph{ndr}) which differs
from the former in several aspects:
\renewcommand{\theenumi}{\alph{enumi}}
\begin{enumerate}
        \item Although $\langle\dot X\rangle$ shows hysteretic behavior
        as a function of $F$ everywhere inside the ordered phase, in the
        \emph{idr} its character is \emph{anomalous} (namely, clockwise)
        whereas in the \emph{ndr} it is \emph{normal} (counterclockwise).
        Moreover, whereas the height of the anomalous hysteresis loop
        increases continuously at the phase boundary in the \emph{idr}
        ($\langle\dot X\rangle$ acts as an order parameter in a
        \emph{second} order phase transition), the disappearance of the
        normal one proceeds by shrinking its width at a more or less
        finite height (the transition at the reentry is of
        \emph{second} order but it is so steep that resembles a
        \emph{first} order one).

        \item The shape of the stationary probability distribution
        function (\emph{pdf}) changes \emph{qualitatively} in going from
        the \emph{idr} to the \emph{ndr} (it becomes \emph{bimodal} and
        remains so as the disordered region is reentered and the
        \emph{pdf} becomes symmetric again, the peak at $\pi$ then being  
        higher than the one at $0$).

        \item Whereas in the \emph{idr} there are \emph{several} solutions
        with $S_m=0$ to the mean-field equations (Eqs.\
        (\ref{eq:15})--(\ref{eq:16})), in the \emph{ndr} there is a
        \emph{unique} solution with $S_m=0$.  Solutions of this kind are
        relevant as a safe starting guess for the Newton-Raphson
        solution of Eqs.\ (\ref{eq:15})--(\ref{eq:16}) in the ordered
        phase, due to the fact that some solutions may suddenly disappear.
\end{enumerate}
Since the transition from anomalous to normal hysteresis in going from
the \emph{idr} to the \emph{ndr} is preceded by the disappearance of a
pair of solutions with $S_m=0$, the line in the phase diagram at which
these disappear (dashed line in Fig.\ \ref{fig:1}) \emph{provides an
estimation} of the place at which the former transition occurs.  Of
course both phenomena are different and so the disappearance of a pair
of solutions with $S_m=0$ \emph{does not imply} an anomalous-to-normal
transition (recall what happens at the left of the dip in the dashed
line of Fig.\ \ref{fig:1}).

Admittedly, all of our results are \emph{mean-field} ones.  Although 
this approximation shows undoubtedly its ability to reveal the 
richness of the phase diagram of this model, it is reassuring to see 
that those of our results that are not original do coincide with 
the numerical simulations for the anomalous hysteresis loop shown in 
Ref.\cite{[6]}.  Nonetheless, the ultimate verification of these 
amazing and potentially useful phenomena lies on the experimentalists' 
side.  We hope to see advances in that direction in a near future.

\bigskip\noindent{\small{\bf ACKNOWLEDGMENTS}}

The authors thank V. Grunfeld for a critical revision of the
manuscript.  Partial support for this work was provided by the
Argentine agencies CONICET (grant PIP 4953/97) and ANPCyT (grant
03-00000-00988).

\section{Appendix}\label{sec.A}

\subsection{The stationary probability distribution function}\label{subsec.A1}

We shall adopt as our standard reference the book by Risken
\cite{[9]}, whose equation (3.67) we have written in the form of Eq.\
(\ref{eq:7}), with $R(x)$ and $S(x)$ defined in Eqs.\
(\ref{eq:8})--(\ref{eq:9}).  At variance with Risken's choice (and in
accord with Ref.\cite{[6]}) we have adopted $q=1$ in our Eq.\
(\ref{eq:2}), equivalent to its Eq.\ (3.68).  Hence Eqs.\ (3.95) are
translated into
\begin{eqnarray}
    D^{(1)}(x)&=&R+\frac{1}{2}\,S\,S',\label{eq:A1}\\
    D^{(2)}(x)&=&\frac{1}{2}\,S^2,\label{eq:A2}
\end{eqnarray}
(the prime indicates a derivative with respect to $x$) whereupon 
$(D^{(2)})'=S\,S'$.

The Fokker-Planck equation (4.46) can be written as
\begin{equation}
    \partial_{t}P(x,t)=-\partial_{x}J(x,t)\label{eq:A3}
\end{equation}
with
\begin{eqnarray}
        J(x,t)&=&D^{(1)}(x)\,P(x,t)-\partial_{x}\left[D^{(2)}(x)\,P(x,t)
        \right]\nonumber\\
        &=&\left(R+\frac{1}{2}\,S\,S'\right)P(x,t)-S\,S'\,P(x,t)-
        \frac{1}{2}\,S^2\,\partial_{x}P(x,t)\nonumber\\
        &=&R\,P(x,t)-\frac{1}{2}\,S\,\partial_{x}\left[S\,P(x,t)\right].
        \label{eq:A4}
\end{eqnarray}
On account of Eq.\ (\ref{eq:A4}), the stationary case of Eq.\
(\ref{eq:A3}) (namely $\partial_{t}P^{st}(x)=0$) implies
$J(x,t)=\mathrm{const.}=\mathit{J}$, whence 
\begin{equation}
    [P^{st}(x)]'=-\frac{2J}{S^2}+\left[\frac{2R}{S^2}-(\ln S)'\right]\,
        P^{st}(x).\label{eq:A5}
\end{equation}
This equation has the form
\begin{equation}
    y'(x)=\alpha(x)+\beta(x)y(x),\label{eq:A6}
\end{equation}
and its general solution is
\begin{equation}
    y(x)=\exp\left[\int_{0}^{x}\beta(x')dx'\right]\,\left\{\int_{0}^{x}
    dx''\alpha(x'')\exp\left[-\int_{0}^{x''}\beta(x')dx'\right]+K
    \right\}.\label{eq:A7}
\end{equation}
As long as $\alpha(x)\neq 0$, the integration constant $K$ can be
chosen so that $y(0)=0$.  Otherwise (in our case for $J=0$) the
solution is
\begin{equation}
        y(x)=K\exp\left[\int_{0}^{x}\beta(x')dx'\right].\label{eq:A8}
\end{equation}

If $\alpha(x+L)=\alpha(x)$ and $\beta(x+L)=\beta(x)$ then
\begin{eqnarray}
    \int_{0}^{x+L}\beta(x')dx'&=&\int_{0}^{L}\beta(x')dx'+
    \int_{L}^{x+L}\beta(x')dx',\label{eq:A9}\\
    \int_{0}^{x+L}dx''\alpha(x'')\exp\left[-\int_{0}^{x''}\beta(x')dx'
    \right]&=&\int_{0}^{x}dx''\alpha(x'')\exp\left[-\int_{0}^{x''}
    \beta(x')dx'\right]\nonumber\\
    &+&\int_{x}^{x+L}dx''\alpha(x'')\exp
    \left[-\int_{0}^{x''}\beta(x')dx'\right].\label{eq:A10}
\end{eqnarray}
The first term in Eq.\ (\ref{eq:A9}) is just a number $N$ and the
second equals $\int_{0}^{x}\beta(x')dx'$ (as can be seen by writing
$x'=x''+L$).  Hence
\begin{equation}
    y(x+L)=e^N\,\left\{y(x)+\exp\left[\int_{0}^{x}\beta(x')dx'\right]\,
    \int_{x}^{x+L}dx''\alpha(x'')\exp\left[-\int_{0}^{x''}\beta(x')dx'
    \right]\right\}.\label{eq:A11}
\end{equation}
By imposing the boundary condition $y(x+L)=y(x)$:
\begin{equation}
    y(x)=\frac{e^N}{1-e^N}\,\exp\left[\int_{0}^{x}\beta(x')dx'\right]
    \,\left\{\int_{x}^{x+L}dx''\alpha(x'')\exp\left[-\int_{0}^{x''}
    \beta(x')dx'\right]\right\}.\label{eq:A12}
\end{equation}

For our case it is $\alpha(x)=-2J/S^2$, $\beta(x)=(2R/S^2)-(\ln S)'$
and $y=P^{st}$ whence, by calling
\begin{equation}
    \phi(x)=-2\int_{0}^{x}\frac{R(x')}{S^2(x')}dx'\label{eq:A13}
\end{equation}
$N=-\phi(L)$ holds, and Eq.\ (\ref{eq:A12}) reads
\begin{equation}
    P^{st}(x)=-\frac{2J\,e^{-\phi(L)}}{1-e^{-\phi(L)}}\,\left\{\frac
        {\exp[-\phi(x)]}{S(x)}\,\int_{x}^{x+L}dy\frac{\exp[\phi(y)]}{S(y)}
        \right\}.\label{eq:A14}
\end{equation}
$J$ is fixed by normalization:
\begin{equation}
    J=\frac{1-e^{\phi(L)}}{2\int_{-L/2}^{L/2}dx\frac{\exp[-\phi(x)]}
    {S(x)}\left\{\int_{x}^{x+L}dy\frac{\exp[\phi(y)]}{S(y)}\right\}}
        =\frac{1-e^{\phi(L)}}{2{\cal N}}.\label{eq:A15}
\end{equation}

\subsection{Contribution to $\phi(L)$ of the interparticle interaction}
\label{subsec.A2}

For $0<A<1/4$, $W(x)$ in Eq.\ (\ref{eq:4}) has a minimum height
$-(1+A)$ at $x=0$ and a maximum height $1-A$ at $x=\pm\pi$.  The
critical points of its derivative are shifted an amount of order $A$
towards $x=0$ from their $A=0$ position of $x=\pm\pi/2$, and so
$S^2(x)$ has its two maxima (of height $1+Q/T$) inside the $\cos x>0$
region.  Now $[S(x)]^{-2}$ has two minima of height $(1+Q/T)^{-1}$ in
that region, whereas in the $\cos x<0$ one it remains of order one.

\subsection{Small-$x$ expansion of $\phi(x)$}\label{subsec.A3}

To first order in $y$, Eqs.\ (\ref{eq:8}) and (\ref{eq:10}) read
$R(y)=-y(1+K_0C_m+4A)+K_0S_m+F$, $S^2(y)=2T[1+(Q/T)(1+4A)^2y^2]$.
Hence in this approximation
\begin{eqnarray}
        T\phi(x)&=&\frac{T(1+4A+K_0C_m)}{2Q(1+4A)^2}\ln[1+(Q/T)(1+4A)^2x^2]
        \nonumber\\
        &-&\sqrt{T/Q}\frac{(K_0S_m+F)}{(1+4A)}\arctan[x\sqrt{Q/T}(1+4A)]
        \nonumber\\
        &\sim&(1+4A+K_0C_m)x^2/2-(K_0S_m+F)x.\label{eq:A21}
\end{eqnarray}
Assuming $K_0S_m=F=0$ we may approximate
\begin{equation}
    P^{st}(x)=\frac{1}{{\cal N}}\,\frac{\exp[-\phi(x)]}{S(x)}
        \sim\frac{1}{\sqrt{2T{\cal N}^2}}\exp\left[-\frac{(1+4A+K_0C_m)x^2}
        {2T}\right]\left[1-\frac{Q}{2T}(1+4A)^2x^2\right]\label{eq:A22}
\end{equation}
which clearly has a maximum at $x=0$. For small $F$ the maximum
shifts toward $(K_0S_m+F)/(1+4A+K_0C_m)$ and that will in turn produce 
a small shift in $S_m$ from zero, in the direction that the maximum
shifts (i.e., that of $F$) whose consequence (by the argument in the
last subsection) is a reversed current.

\subsection{Calculation of the particle current}\label{subsec.A4}

According to Eq.\ (3.85) of Ref.\cite{[9]},
\begin{equation}
        D^{(1)}(x,t)=\left.\lim_{\tau\to 0}\frac{1}{\tau}\langle
        X(t+\tau)-X(t)\rangle\right|_{X(t)=x}=\left.\langle
        \dot{X}(t)\rangle\right|_{X(t)=x}\label{eq:A16}
\end{equation}
(in our case, since $D^{(1)}(x)$ does not depend explicitly on time, this
is the conditional average over realizations of the noise for any $t$).
The unrestricted average at time $t$ is then
\begin{equation}
    \langle\dot{X}(t)\rangle=\int_{-L/2}^{L/2}dx\,D^{(1)}(x)P(x,t)=
        \int_{-L/2}^{L/2}dx\,\left[R+\frac{1}{2}SS'\right]P(x,t).
        \label{eq:A17}
\end{equation}
From Eq.\ (\ref{eq:A4}) it is
\begin{equation}
        \langle\dot{X}(t)\rangle=\int_{-L/2}^{L/2}dx\,J(x,t)+
        \left[D^{(2)}(x)P(x,t)\right]_{-L/2}^{L/2},\label{eq:A18}
\end{equation}
so for \emph{periodic} $P(x,t)$ it is
\begin{equation}
        \langle\dot{X}(t)\rangle=\int_{-L/2}^{L/2}dx\,J(x,t)\label{eq:A19}
\end{equation}
and in the stationary state, where $J(x,t)=\mathrm{const.}=\mathit{J}$:
\begin{equation}
    \langle\dot X\rangle=JL.\label{eq:A20}
\end{equation}

\begin{figure}
\caption{Phase diagram of the model for $T=2$, $A=0.15$ and $F=0$.
The ordered region lies above the full line.  Above the dashed line
there may exist up to three solutions when $S_m\neq 0$, whereas below
it there may exist at most one.  The squares represent states at which
the shape of $P^{st}(x)$ and the behavior of $\langle\dot X\rangle$ as
a function of $F$ have been investigated.  They correspond to $K_0=10$
and $Q=$1, 3, 6, 9, 12, 16 and 21 respectively.}
\label{fig:1}
\end{figure}

\begin{figure}
\caption{Illustration of the passage from the \emph{idr} to the
\emph{ndr}, as the noise intensity $Q$ increases for $K_0=10$ (solid
line: $Q=5.95$; dashed line: $Q=6.5$).  Only the solution with $C_m<0$
survives after the disappearance of the other two.}
\label{fig:2}
\end{figure}

\begin{figure}
\caption{$P^{st}(x)$ for $K_0=10$ and the values of $Q$ in Fig.\ 1.
For $Q=1$ and $Q=21$ (solid line) it is symmetric, being asymmetric
for the remaining values (dashed lines).  For $Q$ between 6 and 9 it
becomes bimodal and as $Q$ increases the peak with larger $|x|$
overtakes the other one, although it never reaches beyond 0.5}
\label{fig:3}
\end{figure}

\begin{figure}
\caption{The order parameter $V_m=\langle\dot X\rangle$ (particle 
current) as a function of $Q$ for $K_0=10$ and $F=$0 (squares), 0.3 
(upward triangles) and 0.44 (downward triangles).  The vertical thick 
line signals the transition from anomalous to normal hysteresis.}
\label{fig:4}
\end{figure}

\begin{figure}
\caption{(a) $V_m=\langle\dot X\rangle$ vs $F=$ for $K_0=10$ and 
$Q=5.97$ (just on the left of the dashed line of Fig.\ 1).  The stable 
solutions are those with larger $V_m$ values; the other two solutions 
lie on the branches of the $C_m>0$ and $C_m<0$ ones for $S_m=0$ (the 
solutions with $C_m\approx 0$ are not included).  (b) Same as for 
$Q=6.0$: one of the unstable solutions has receded from the $F\approx 
0$ region (together with the $C_m\approx 0$ one, not shown).  (c) Same
as for $Q=6.1$, showing a complete recession from
the $F\approx 0$ region.  (d) Same as for $Q=6.5$: not until the dotted 
line has completely disappeared do solutions in the stable branch 
appear for $|F|>0.5$ and normal hysteresis sets in.}
\label{fig:5}
\end{figure}

\begin{figure}
\caption{$V_m=\langle\dot X\rangle$ vs $F$ for $K_0=10$ and $Q=$1, 
1.7, 3, illustrating the appearance of multiple solutions and of
anomalous hysteresis as the ordered region is reached from the left.}
\label{fig:6}
\end{figure}

\begin{figure}
\caption{$V_m=\langle\dot X\rangle$ vs $F$ for $K_0=10$ and $Q=$9, 12, 
16, 21, illustrating the disappearance of the normal hysteresis loop
at the rentrance (the disordered region is reached from the left).}
\label{fig:7}
\end{figure}

\begin{figure}
\caption{(a) $V_m=\langle\dot X\rangle$ vs $F=$ for $Q=6.0$ and 
$K_0=$7.25, 8.0 and 10.0, illustrating the way the transition from 
normal to anomalous hysteresis proceeds as $K_0$ increases at 
$Q=\mathrm{const.}$ above the dip in Fig.\ 1.  (b) Same as for
$K_0=$14.0 and 18.0: the cusp at the endpoints of the anomalous loop 
has developed into a curl, thus reducing further its excursion.}
\label{fig:8}
\end{figure}

\begin{figure}
\caption{$V_m=\langle\dot X\rangle$ vs $F=$ for $Q=9.5$ and $K_0=$16.0 
(\emph{idr}) and 13.0 (\emph{ndr}).  For $K_0=13.0$ there is still a 
remnant of the curl existing in the anomalous zone.}
\label{fig:9}
\end{figure}

\begin{figure}
\caption{Illustration of the passage from \emph{idr} to 
\emph{ndr}, as the coupling $K_0$ decreases for $Q=10$.  As in Fig.\ 
2, only the solution with $C_m<0$ survives after the disappearance of 
the other two.}
\label{fig:10}
\end{figure}

\begin{figure}
\caption{$V_m=\langle\dot X\rangle$ vs $F=$ for $Q=4.0$ (at the left
of the dip in Fig.\ 1) and $K_0=$9.0 (\emph{idr}) and 8.2 (\emph{ndr}).
One of the branches of unstable solutions has disappeared (together
with the $C_m\approx 0$ one, not shown), but the hysteresis loop
remains anomalous.}
\label{fig:11}
\end{figure}


\begin{references}

\bibitem{[1]} R. P. Feynman, R. B. Leighton, and M. Sands, \emph{The
Feynman Lectures on Physics, Mainly Mechanics, Radiation, and Heat,
Volume I}, Addison-Wesley (1963), chapter 46.

\bibitem{[3]} R. D. Vale and F. Oosawa, Adv.\ Biophys.\ {\bf 26}, 97
(1990); A. Ajdari and J. Prost, C. R. Acad.\ Sci.\ Paris {\bf 315},
1635 (1992).

\bibitem{[4]} M. O. Magnasco, Phys.\ Rev.\ Lett.\ {\bf 71}, 1477
(1993); R. D. Astumian and M. Bier, Phys.\ Rev.\ Lett.\ {\bf 72},
1766 (1994); C. R. Doering, W. Horsthemke and J. Riordan, Phys.\
Rev.\ Lett.\ {\bf 72}, 2984 (1994); R. Bartussek, P. H\"anggi and J.
G. Kissner, Europhys.\ Lett.\ {\bf 28}, 459 (1994).

\bibitem{[5]} P. Reimann, Phys.\ Rep.\ {\bf 290}, 149 (1997).

\bibitem{[6]} P. Reimann, R. Kawai, C. Van den Broeck and P. H\"anggi,
Europhys.\ Lett.\ {\bf 45}, 545 (1999).

\bibitem{[8]} S. Mangioni, R. Deza, H. S. Wio and R. Toral, Phys.\
Rev.\ Lett.\ {\bf 79}, 2389 (1997); S. Mangioni, R. Deza, R. Toral and
H. S. Wio, Phys.\ Rev.\ E {\bf 61}, 223 (2000).

\bibitem{[7]} C. Van den Broeck, P. Reimann, R. Kawai, and P.
H\"anggi, XV Sitges Euroconference on ``Statistical Mechanics of
Biocomplexity'' (1998); P. Reimann, C. Van den Broeck and R. Kawai,
Phys.\ Rev.\ E {\bf 60}, 6402 (1999); J. Buceta, J. M. R. Parrondo, C.
Van den Broeck, and J. de la Rubia, Phys.\ Rev.\ E {\bf 61}, 6287
(2000).

\bibitem{[10]} http://www.kawai.phy.uab.edu/research/motor/bm.mov.

\bibitem{[9]} H. Risken, {\em The Fokker-Planck Equation, Methods of
Solution and Applications}, 2nd edition, Springer-Verlag
(Berlin, 1989).

\end{references}
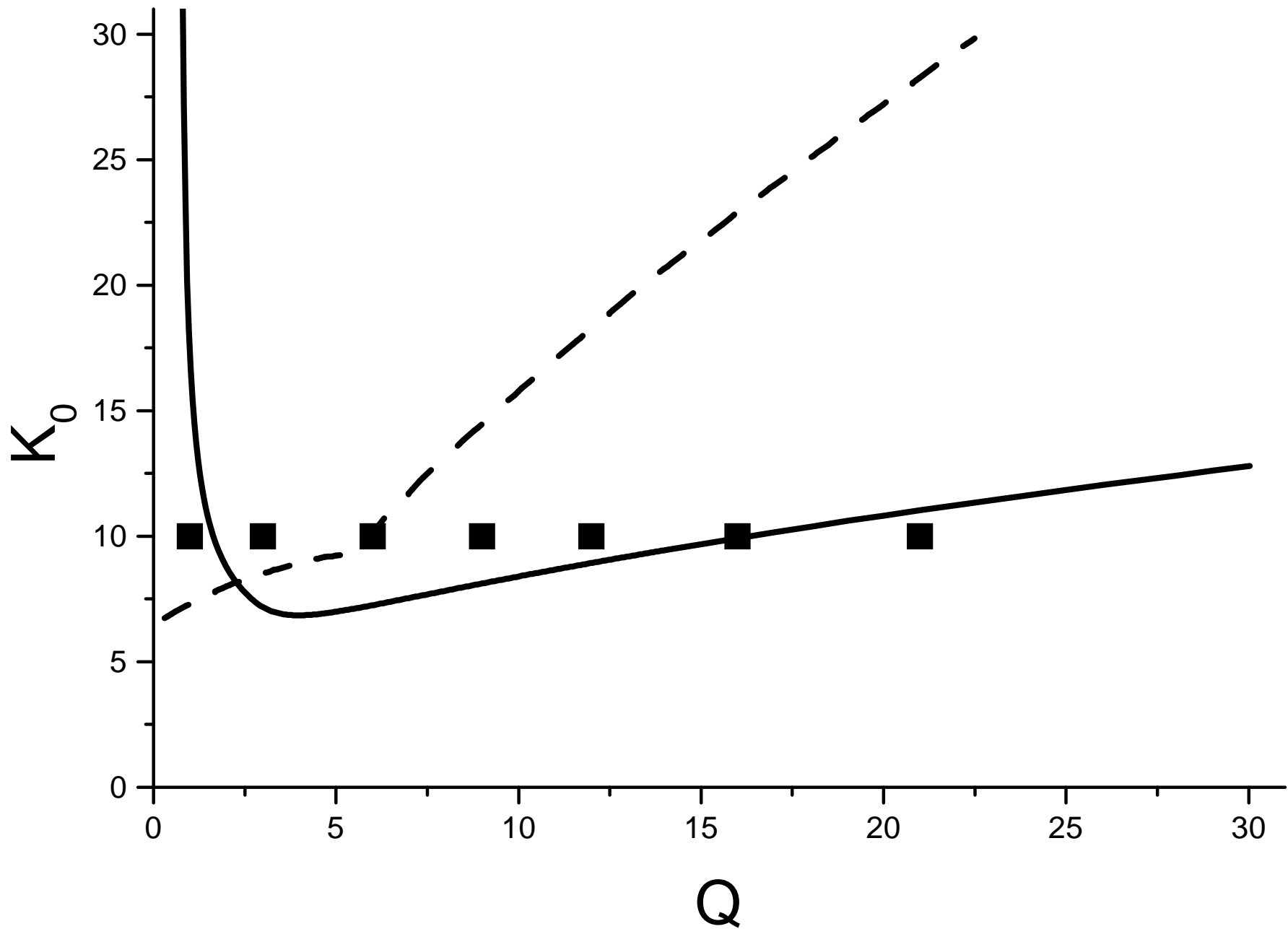
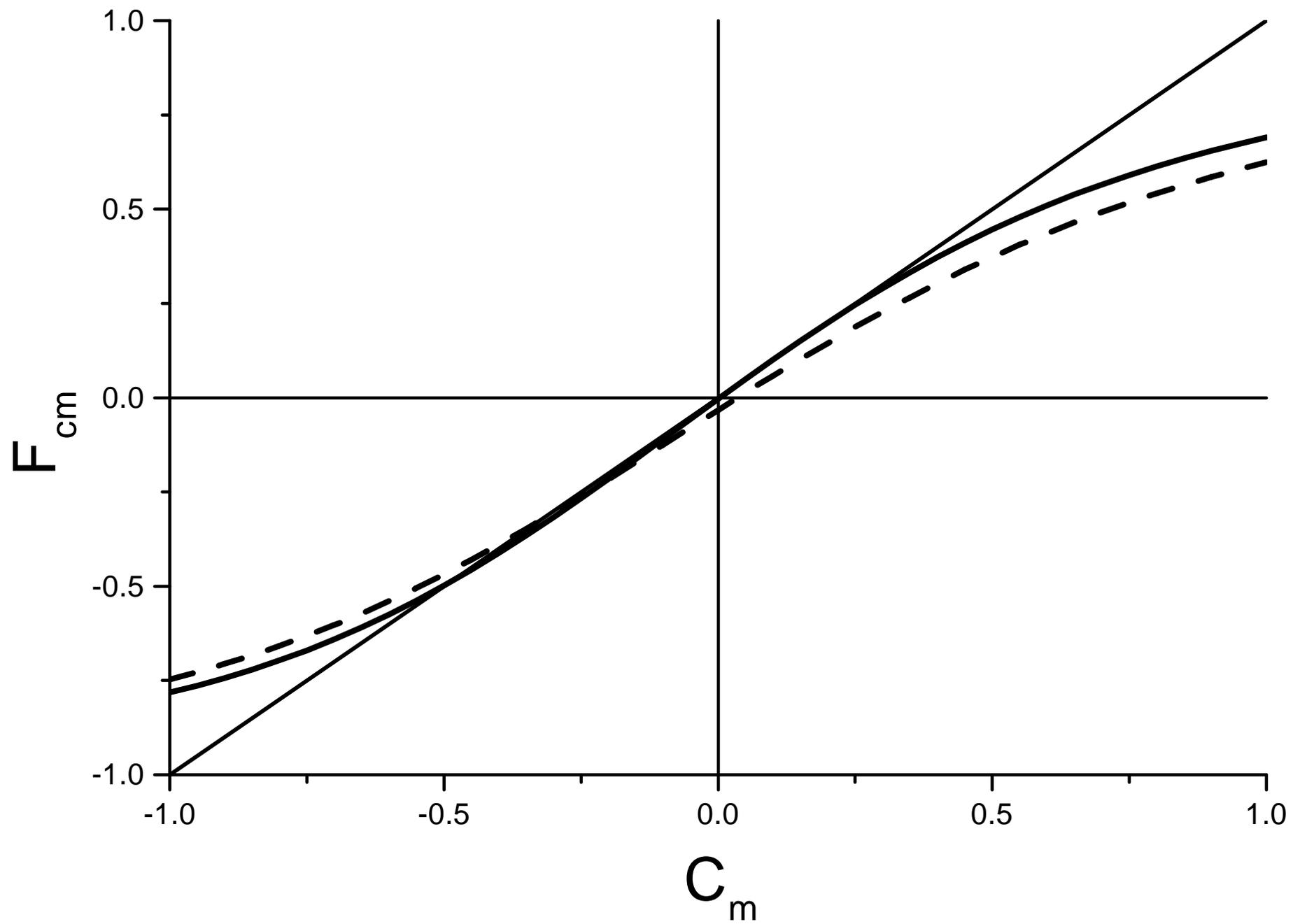
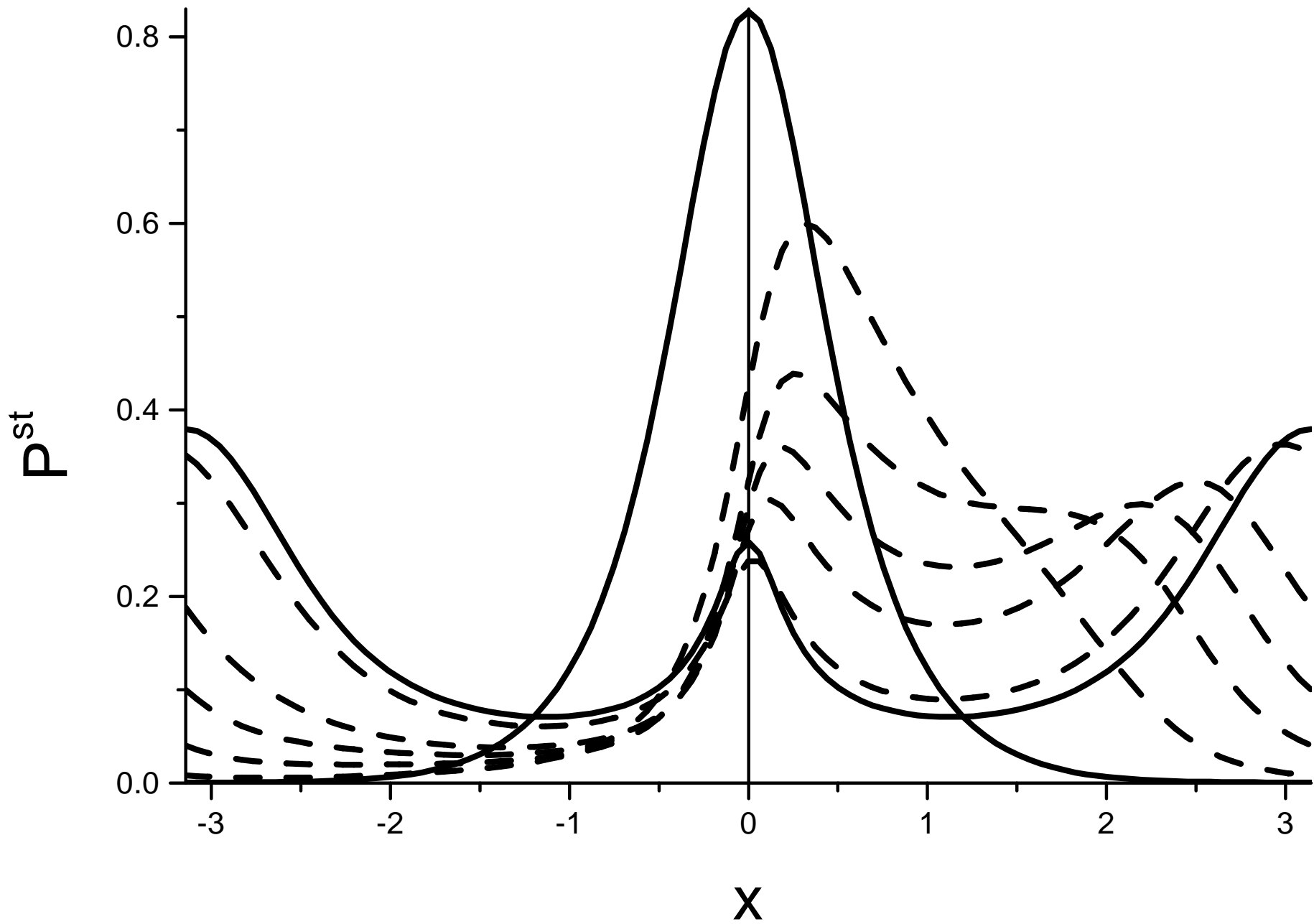
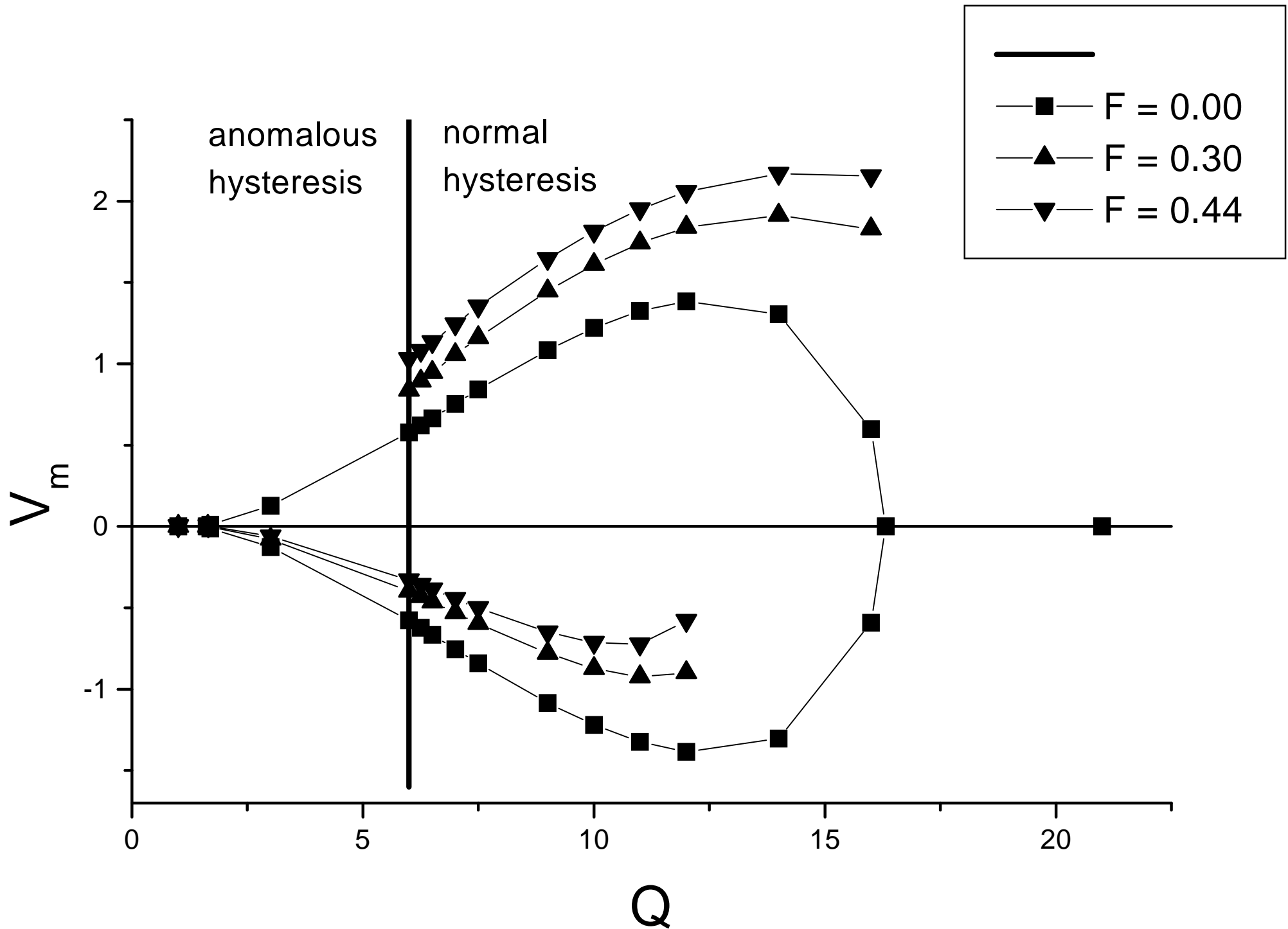
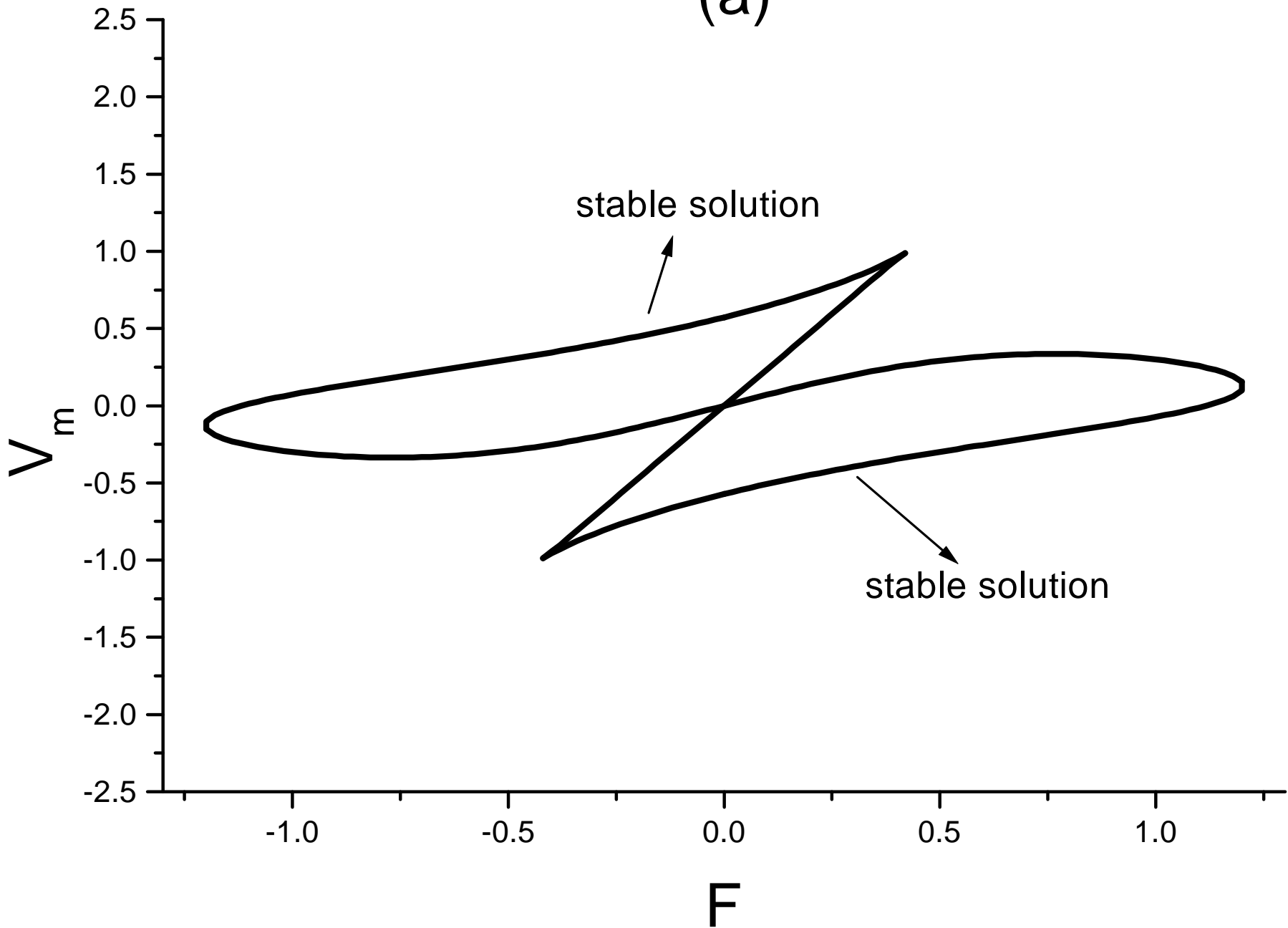
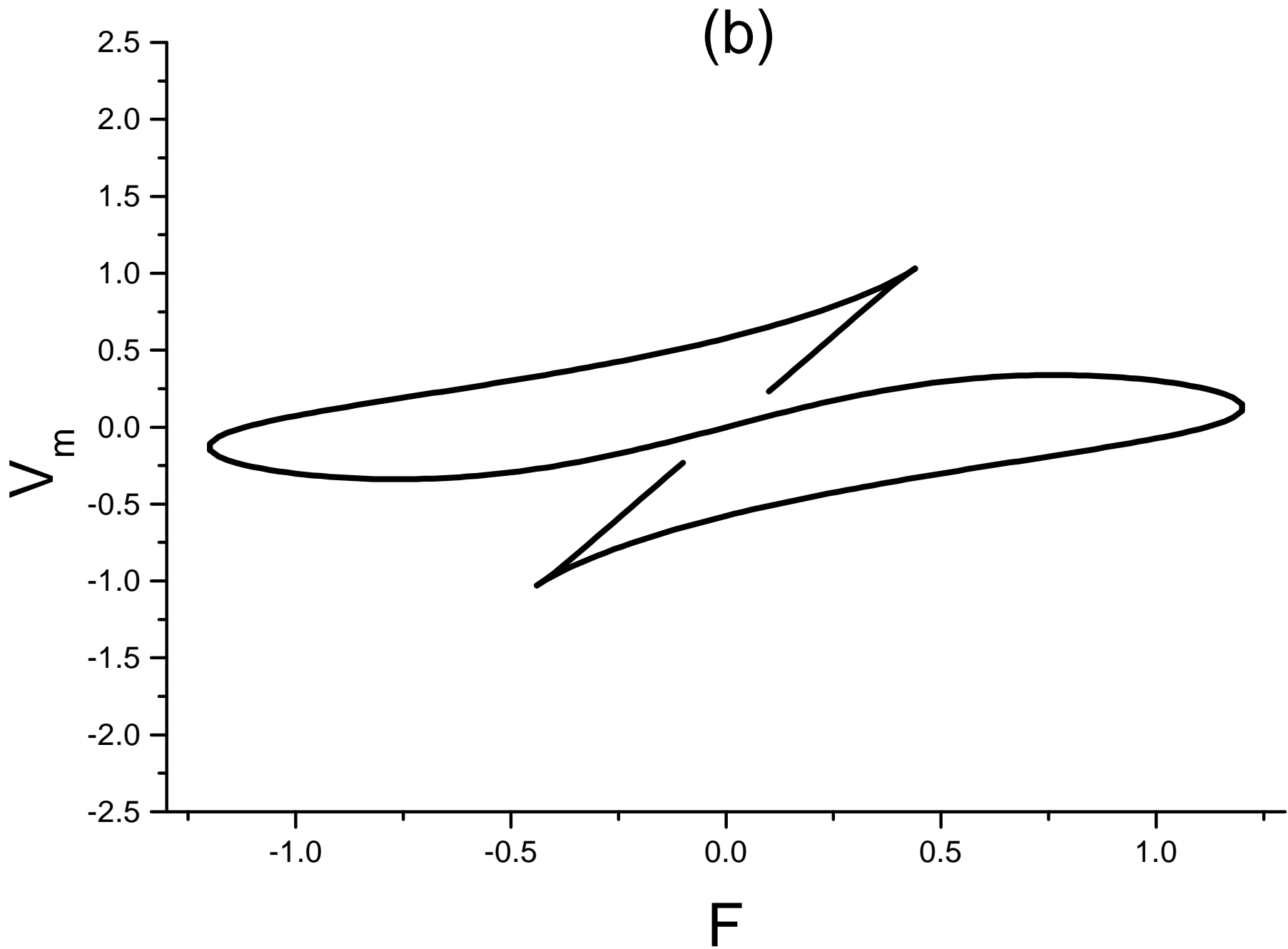
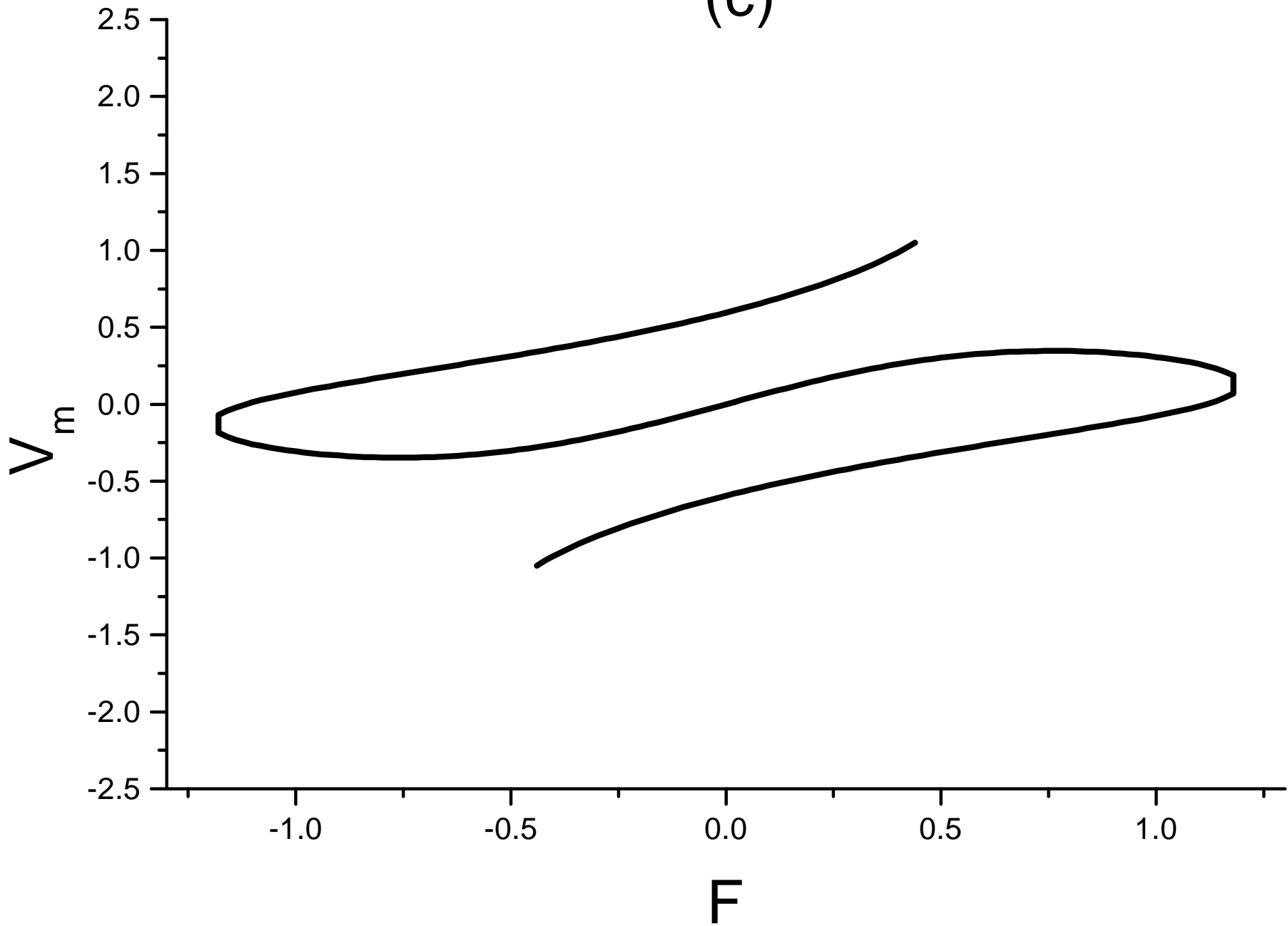
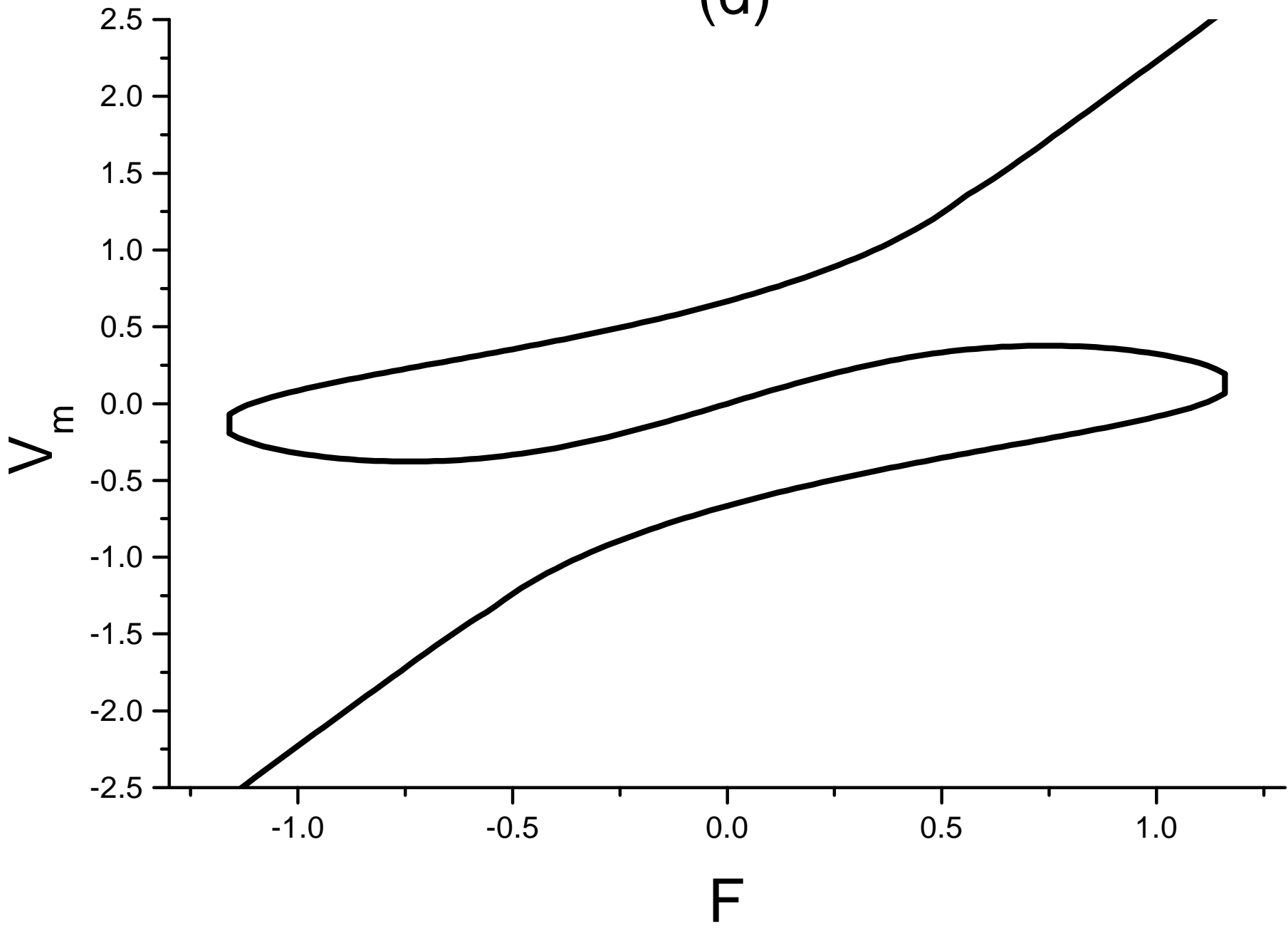
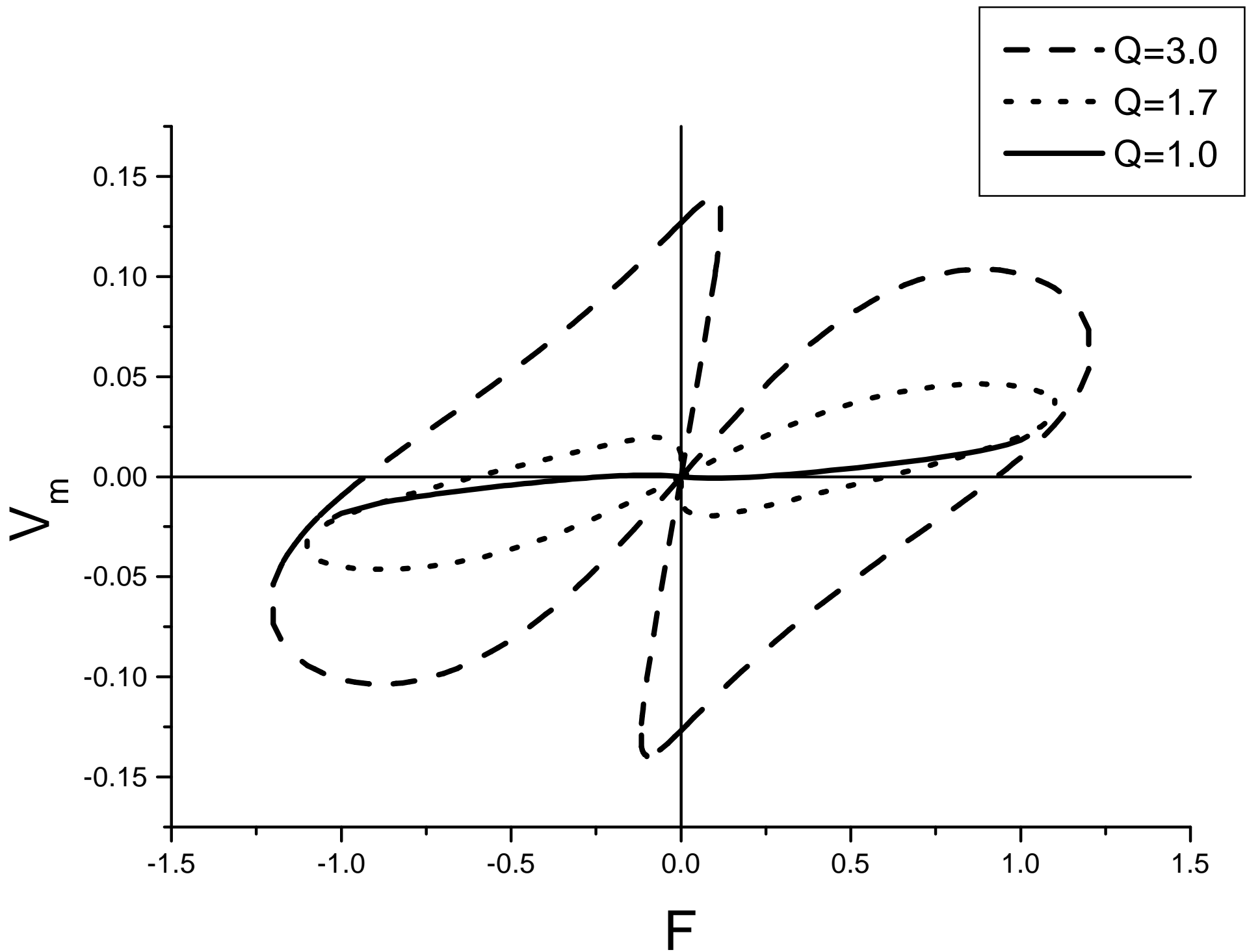
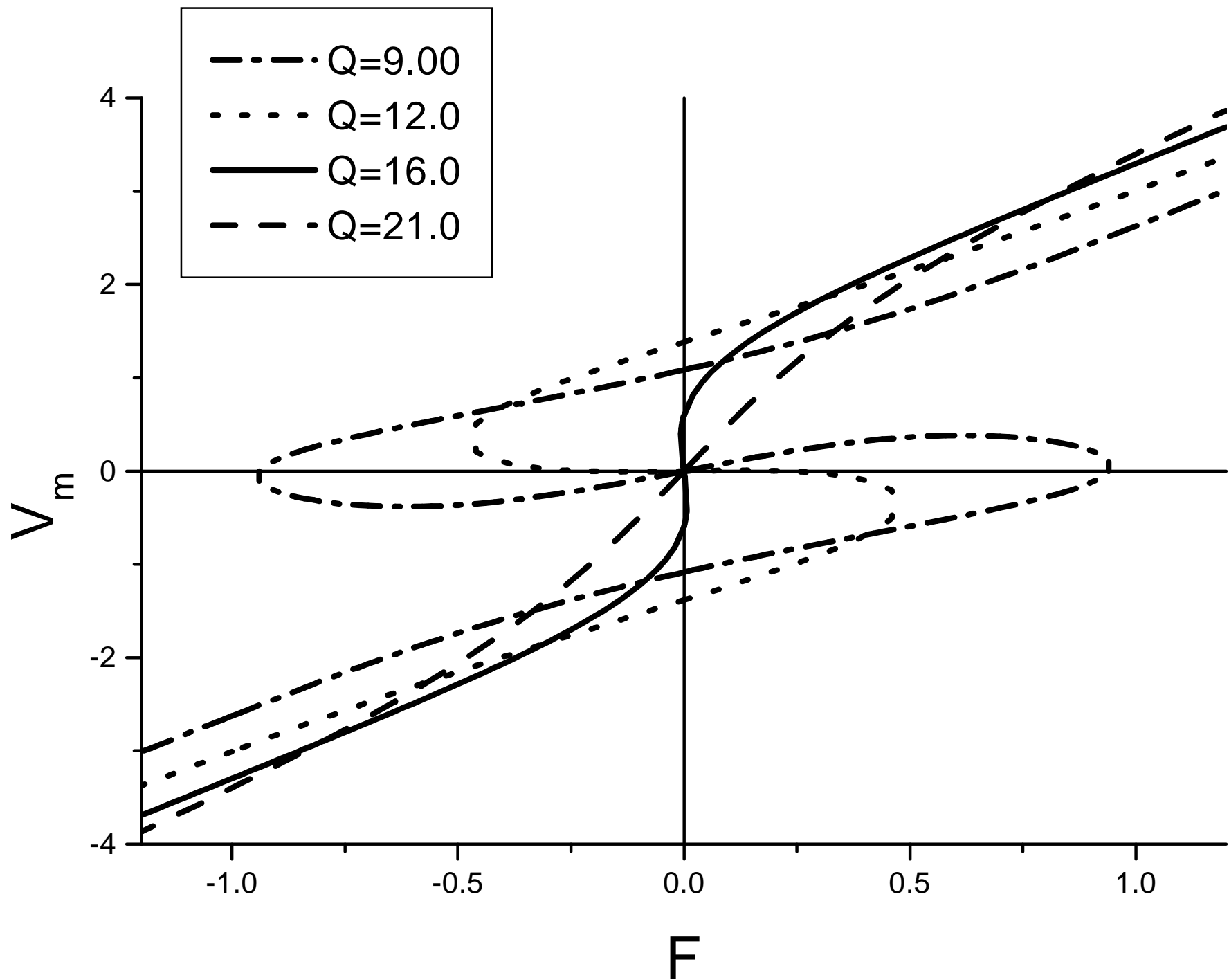
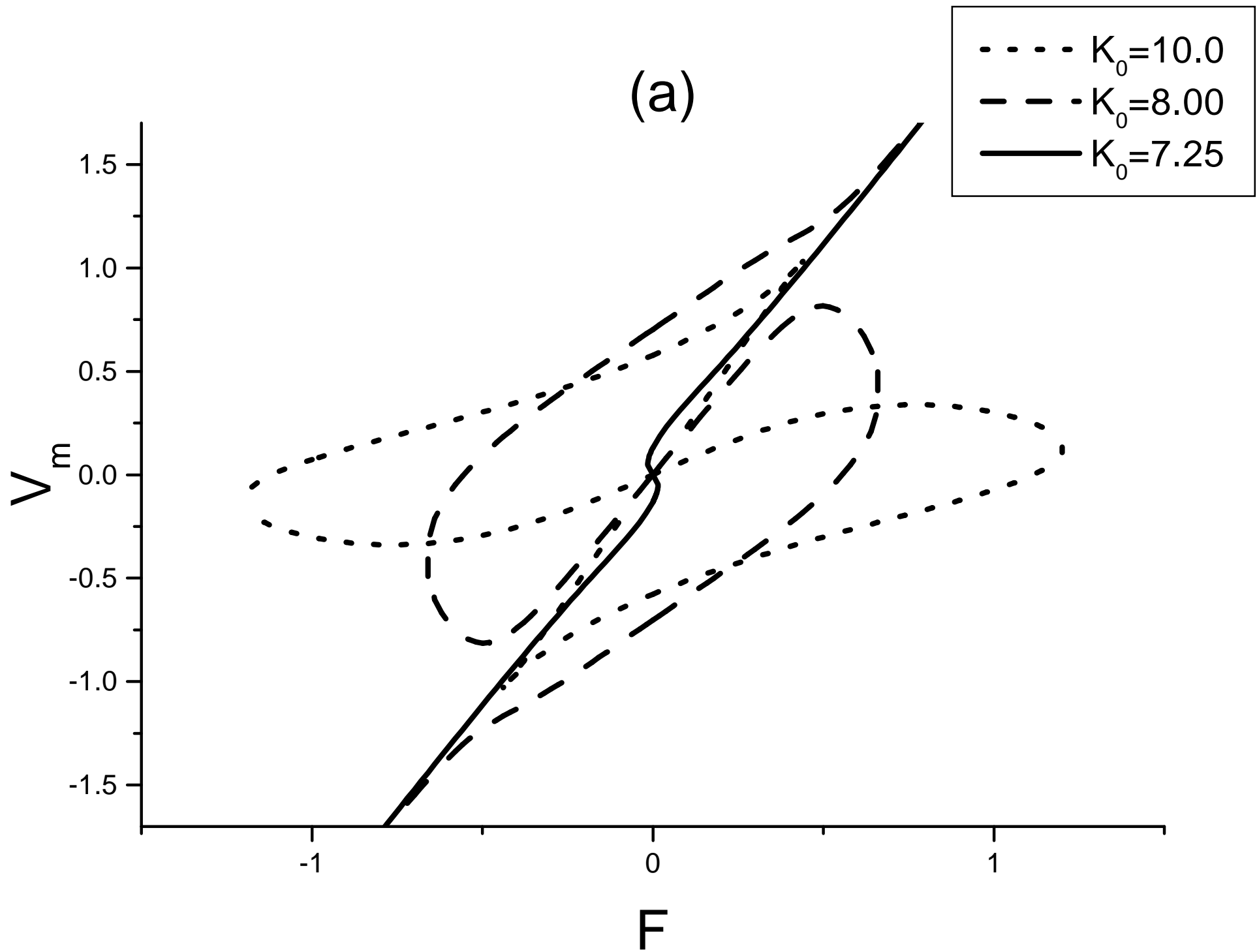
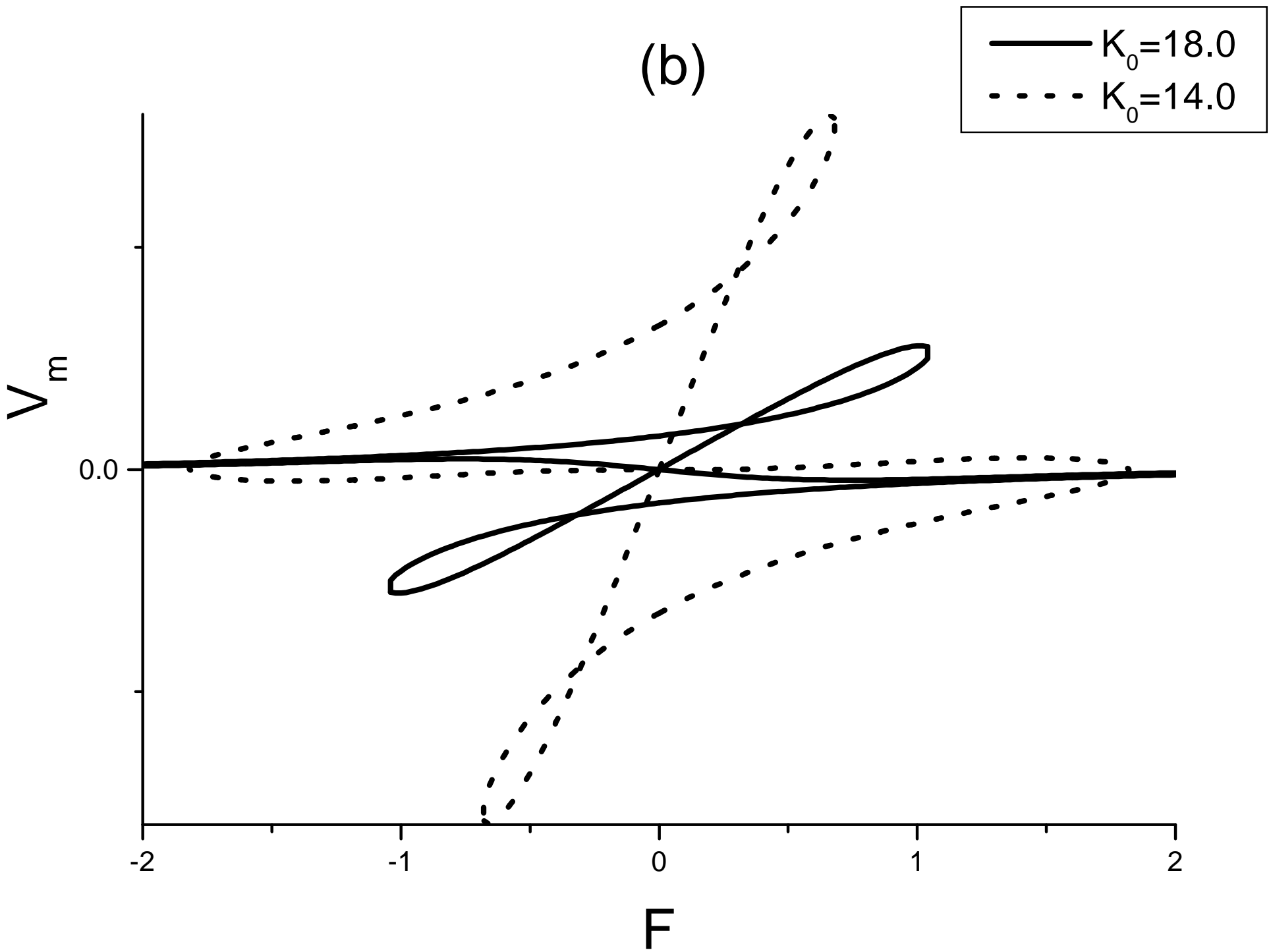
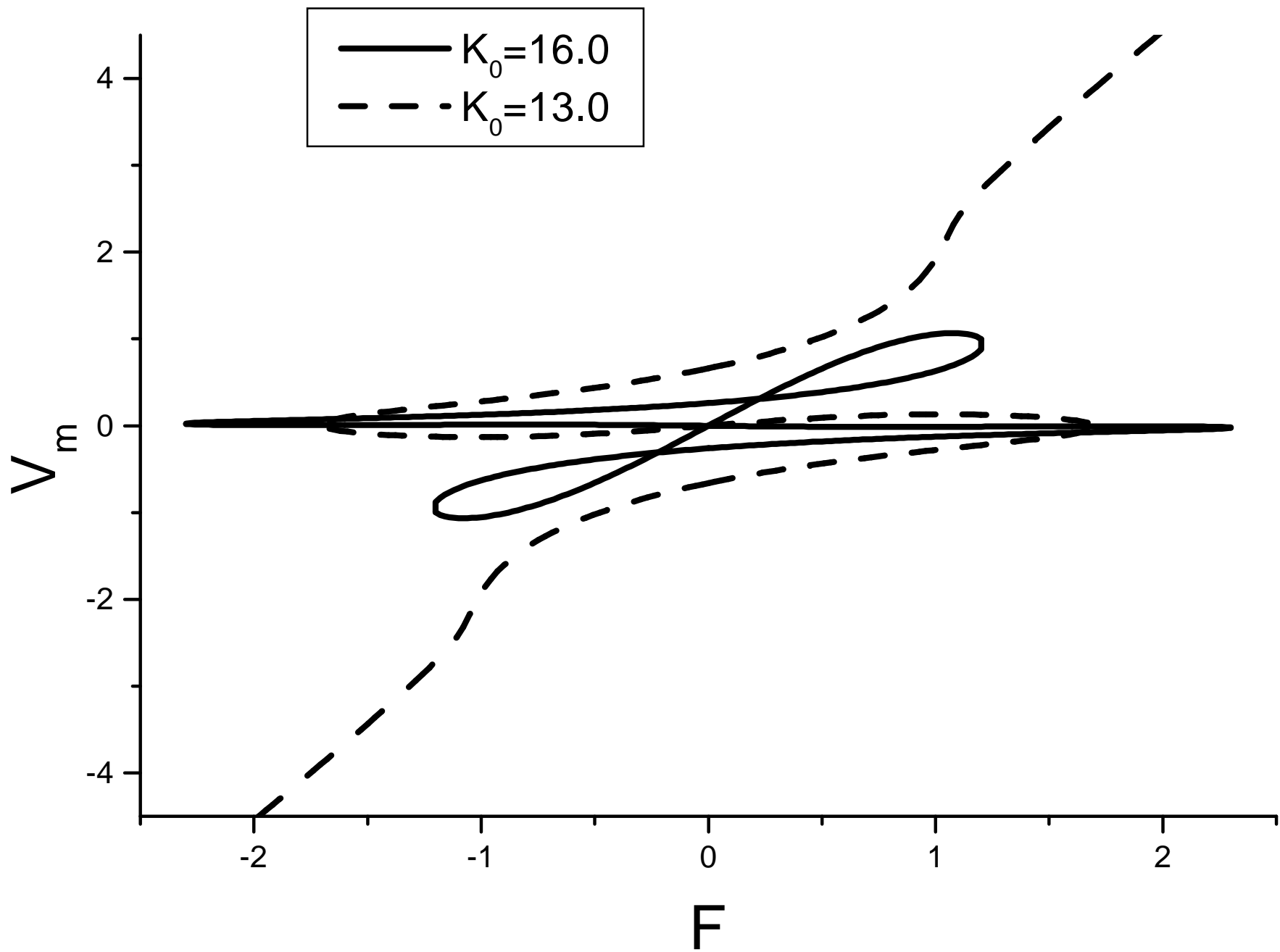
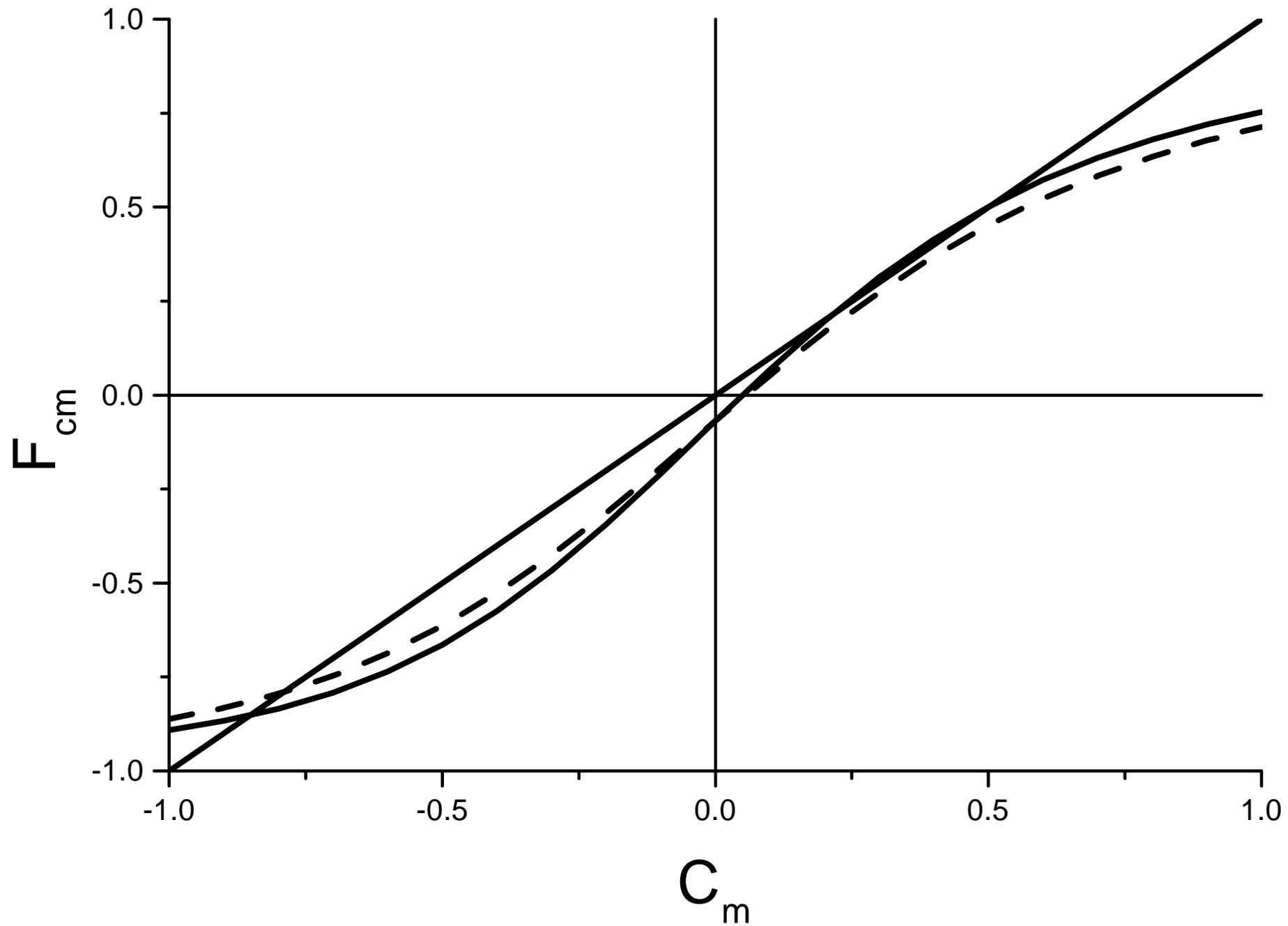
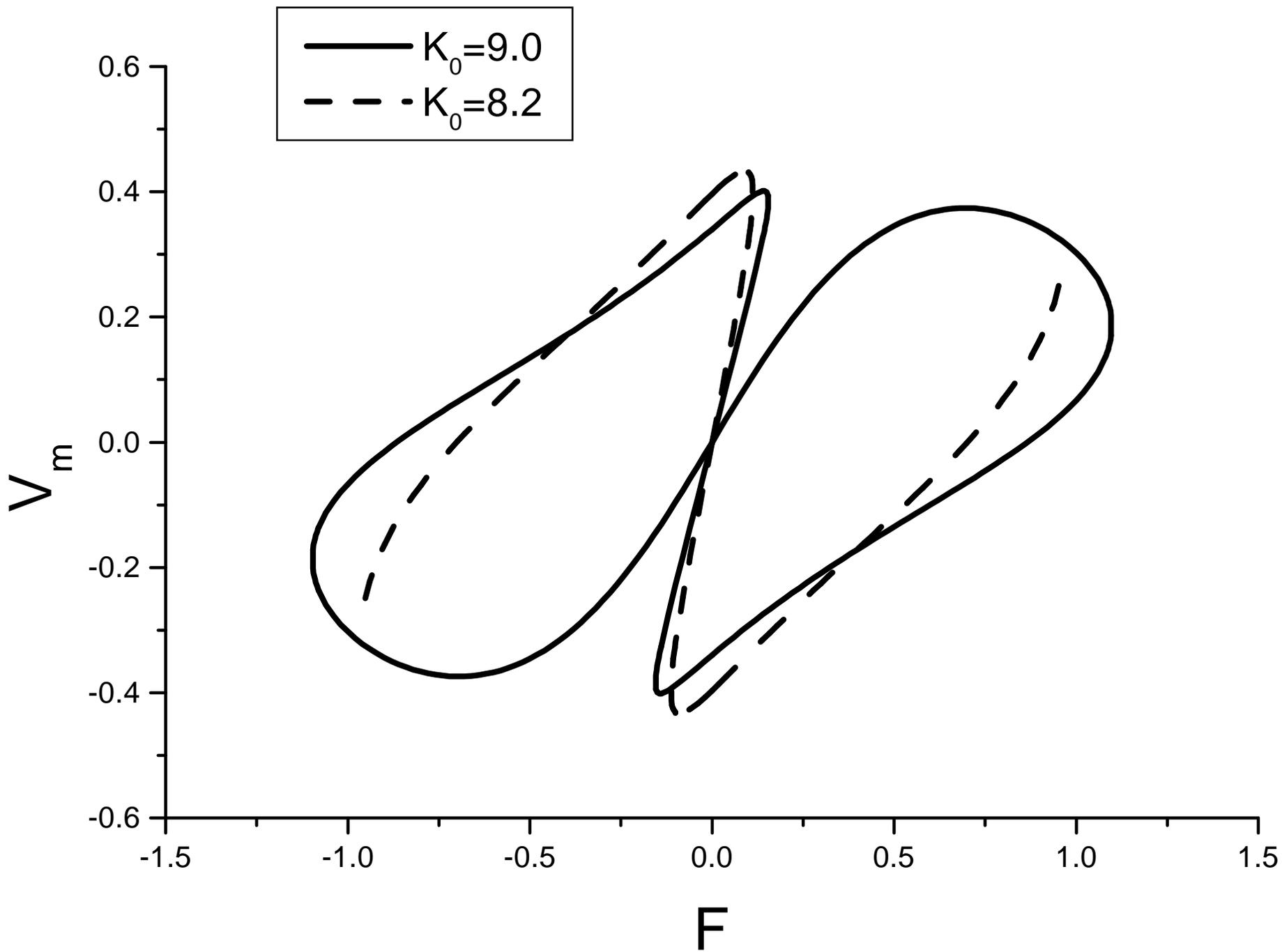
\end{document}